\documentclass[a4paper,12pt]{article}
\usepackage[dvips]{graphicx,color,epsfig,rotating}
\let\section=\subsection     \let\subsection=\subsubsection                

\begin{document}
\begin{center}
{\large \bf HADRONIC IN-MEDIUM EFFECTS} \\[2mm]
{\large \bf WITH ELEMENTARY PROBES} \\[5mm]
U. MOSEL\footnote{http://theorie.physik.uni-giessen.de}\\[5mm]
{\small \it Institut fuer Theoretische Physik, Universitaet
Giessen\\ D-35392 Giessen, Germany\\[8mm]}
\end{center}

\begin{abstract} \noindent The sensitivity of dilepton production in elementary
reactions of photons and pions with nuclei on in-medium changes of
hadronic properties is studied. It is shown that this sensitivity is
comparable to that encountered in ultrarelativistic heavy-ion
collisions. It is also shown how a significant broadening of the vector
mesons affects other photonuclear processes.
\end{abstract}

\section{Introduction}

Results of the CERES experiment show a significantly higher dilepton
yield in the invariant mass region between about 200 and 600 MeV
compared to that from a hadronic cocktail based on known reaction
channels. After adding in the -- expected -- secondary $\pi \pi
\rightarrow \rho$ yield the observed yield is still underestimated by a
hadronic cocktail by a 'sensitivity factor' of about 3. While there is
also an attempt to explain these observations in terms of standard
'classical' hadronic sources \cite{Koch} by exploiting their
experimental uncertainties, most theoretical explanations invoke a
change of the properties of the vector meson masses and widths. Some of
these are based on 'classical' collision broadening whereas others
involve $\rho-a_1$ mixing as a precursor to chiral symmetry restoration
\cite{Rapp}. However, all these attempts have in common that in a first
step they determine the in-medium properties in hot and dense
equilibrium and then use these locally either in transport calculations
or in more simplified expansion scenarios.

The latter procedure is obviously not a-priori correct for an
ultrarelativistic heavy-ion collision which -- at least in its early
stages -- proceeds far away from equilibrium. It is worthwhile to
realize that even in ultrarelativistic heavy-ion collisions with their
high peak densities about 60 \% of all dileptons are produced at
densities below $2 \rho_0$ (see Fig.\ 7 in \cite{Cassing00}). This is
so because the observed yield inherently contains an integration over
the whole reaction history and thus the late stages of the reaction
with high pion densities contribute via $\pi^+ + \pi^- \rightarrow e^+
+ e^-$ a significant part to the total observed dilepton yield.

It is, therefore, intriguing to ask if other reactions involving more
elementary probes on nuclei that proceed closer to equilibrium can be
used to investigate the question of in-medium changes of hadronic
properties. While the density in such reactions obviously is always
below $\rho_0$, this disadvantage may be overcome by the 'cleaner' and
more stable environment in which dilepton production proceeds in such
reactions.

In this talk I show that indeed pion- and photon-induced dilepton
production on nuclei is  as sensitive to possible in-medium changes as
are heavy-ion reactions. I will also -- in the last part of this talk
-- show how in-medium changes of vector mesons are expected to show up
not only in dilepton spectra, but also in photo-absorption processes.
Such a connection, if it can be firmly established, clearly adds to the
consistency of our picture of in-medium changes of hadrons.

Details to all the points I am going to discuss in this talk can be
found in a number of recent publications \cite{Effepi,Effegam,Lehr} and
in particular in the PhD thesis of Martin Effenberger \cite{EffePhD}.

\section{Model}

The calculations shown in this talk are all based on a new development
in transport theory that allows one to also tranport broad resonances.
While former calculations (for a review see \cite{B-C}) had to employ
simplifiying assumptions the calculations of Effenberger et al.\
\cite{Effegam}, for the first time, implemented a transport-theoretical
treatment of broad mesons and treated the widths of the vector mesons
consistently. The theoretical breakthrough here was to transport not
the phase-space density $f(\vec{x},\vec{p},t)$, but a 'spectral phase
space density' $F(\vec{x},\vec{p},\mu,t)$ that contains also
information about the spectral function of the resonance.

In \cite{Effegam} (see also \cite{Effeoff}) we have discussed in detail
that a consistent formulation of the scattering term, that takes the
spectral function of the particles into account, puts automatically
intrinsically broad, collision-broadened resonances (such as, e.g.\ the
$\rho$ meson) -- because of their frequent decay and reformation as
they traverse the nucleus -- back on their free spectral function when
they leave the nucleus. A problem arises, however, when particles with
intrinsically sharp spectral functions, such as the nucleon or the
$\omega$ meson, get collision broadened. In such a case the repetitive
decay and reformation is not frequent enough so that these particles
would emerge from the collision still collision-broadened. In order to
surpress this unphysical behavior we introduced an ad-hoc potential
that drives all particles back on mass-shell when they leave the
nucleus \cite{Effegam}. It is amusing that this potential, that we
intuitively guessed in ref.\ \cite{Effegam}, can  -- for sharp
resonances, where $\Gamma_{tot} \approx \Gamma_{coll}$ -- actually be
derived from the Kadanoff-Baym equations \cite{Leupold,CJ}.

The results presented here are based on this method (for more details
see \cite{Effegam,EffePhD,Effeoff}). The equation of motion for the
spectral phase space distribution reads
\begin{equation}
\left(\frac{\partial}{\partial t} + \vec{\nabla}_p H_i \vec{\nabla_x} -
      \vec{\nabla}_x H_i \vec{\nabla}_p \right) F_i
      = G_i A_i - L_i F_i
\end{equation}
with the spectral phase-space-density
\begin{equation}
F_i(\left(\vec{x},\vec{p},\mu,t \right) = A_i\left(\vec{x},\vec{p},\mu,t
\right) f(\vec{x},\vec{p},\mu,t ) ~,
\end{equation}
the spectral function
\begin{equation}
A_i(\mu) = \frac{2}{\pi} \frac{\mu^2 \Gamma_{\rm tot}(\mu)}{\left(\mu^2
            - M_i^2\right)^2 + \mu^2 \Gamma^2_{\rm tot}(\mu)} ~,
\end{equation}
and the usual phase-space density $f(\vec{x},\vec{p},\mu,t)$. As
stressed in particular by Knoll \cite{Knoll} it is essential to
maintain the consistency condition that the total width appearing in
the spectral function is related to the loss-term $L$ in the transport
equation
\begin{equation}
\Gamma_{\rm tot} = \gamma L ~.
\end{equation}
Since this condition presents a major consistency problem it has been
fulfilled in the calculations reported here only for the vector mesons.
In \cite{Effeabs} it was shown that the two-body collisional width does
not broaden the nucleon resonances significantly. For all further
details I refer the reader to \cite{Effegam,Effeoff,EffePhD}.

\section{Pion-Induced Dileptons}

With the availability of a $\pi^-$ beam at GSI and the detectorsystem
HADES \cite{Friese} it will be possible to explore dilepton production
in pion-induced reactions on nuclei. We have, therefore, performed
extensive studies of such reactions \cite{Effepi}. Input to these
calculations are cross sections predicted by the Manley coupled channel
analysis of $\pi N$ data \cite{Manley} on which our analysis is based.

The transport calculations allow one to follow the collision history of
all vector mesons produced. With realistic $VN$ cross sections we
obtain a collision width $\Gamma_{\rm coll} = \rho \sigma_{VN} v_{VN}$
of up to 500 MeV for the $\rho$ meson and 70 MeV for the $\omega$, both
at normal nuclear matter density \cite{Effegam}.

\begin{figure}[ht!]
\centerline{\psfig{figure=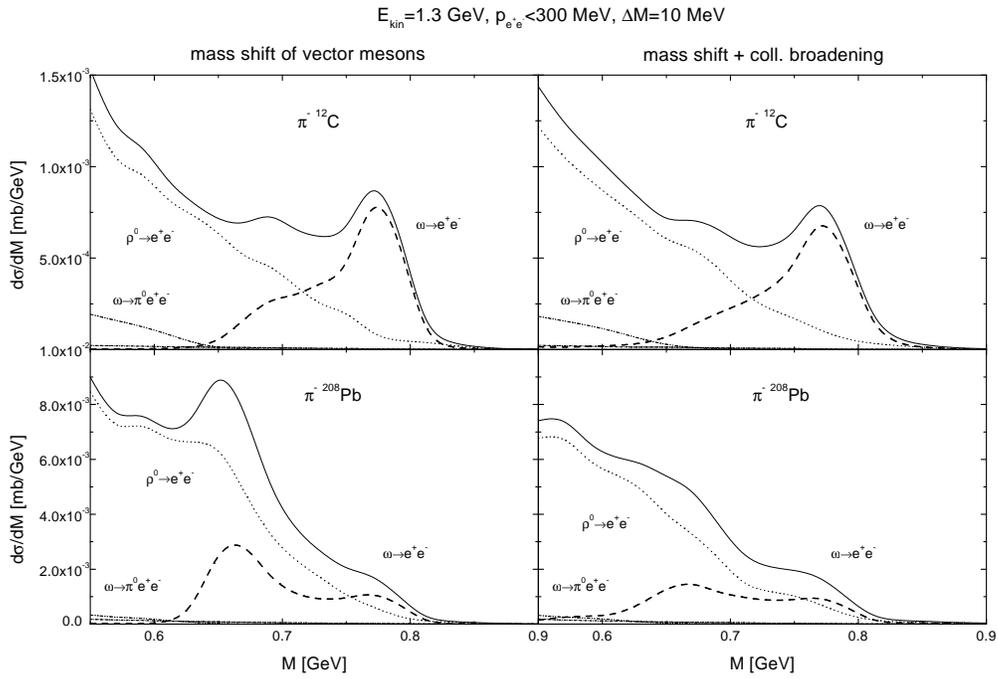,width=14cm}} \caption{\small
Effects of medium modifications on dilepton invariant mass spectra in
$\pi^- C$ and $\pi^- Pb$. Here only dileptons with total momentum $<$
300 MeV haven been considered} \label{picut}
\end{figure}

The overall sensitivity to these effects is about a factor 2, which is
close to the expected systematic uncertainty of these transport
calculations. However, this sensitivity can be enhanced by performing
appropriate cuts on the dilepton cm momenta. Fig.\ \ref{picut} shows
that with a cut that selects only slow vector mesons the sensitivity to
various in-medium scenarios, in particular for the $\omega$ meson, is
clearly strong enough to be seen experimentally.

\section{Photon-Induced Dileptons}

\begin{figure}[ht!]
\centerline{\psfig{figure=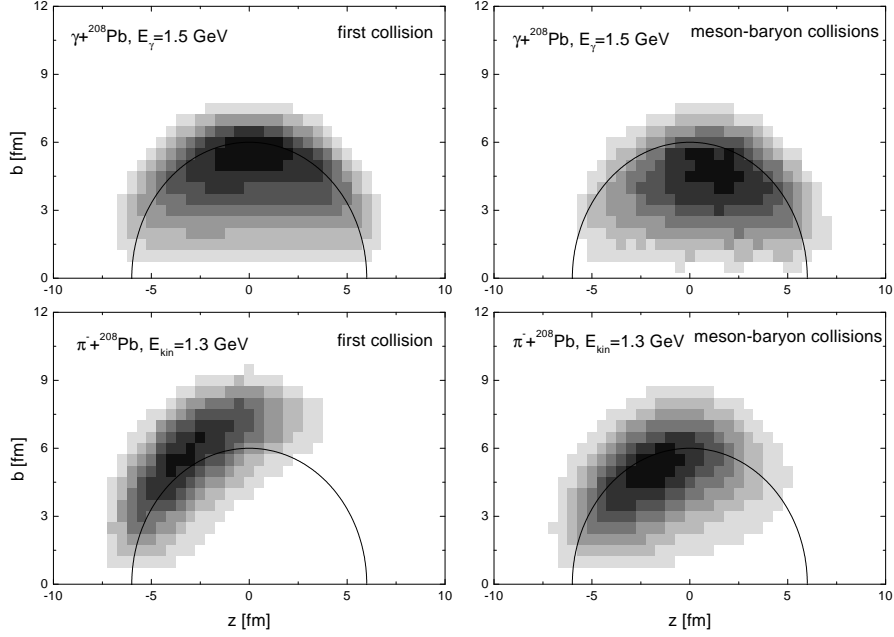,width=13cm}} \caption{\small
Spatial distribution of first collisions (left) and subsequent
meson-baryon collisions (right) for $\gamma+Pb$ (top) and $\pi + Pb$
(bottom)} \label{pi-gamma ill}
\end{figure}

Pions have one possible disadvantage for such in-medium studies: they
experience a strong initial-state interaction and thus a pion-beam
illuminates only a part of the nucleus. This is visible in Fig.\
\ref{pi-gamma ill} which shows in the two lower figures the location of
the first collision of the pion with the nucleons of the target (left)
and all meson-baryon collisions (right).
The upper two figures give the same information for a photon beam. The
left part shows clearly how the photon illuminates the whole nuclear
volume and not just the surface. Subsequent collisions remember this
initial state as shown in the right part of this figure. Thus, we
expect that photon-induced experiments will be even more sensitive to
in-medium changes of hadronic properties than experiments with pions.

That the theory, with the proper, consistent treatment of in-medium
broadening describes the 'normal' photonuclear processes very well is
shown in a comparison with data for photo-pion, photo-2pion and
photon-eta production data at energies up to 800 MeV \cite{Lehr}. We
are thus quite confident that we have these photonuclear processes well
under control and can thus use these methods to predict and investigate
the sensitivity of photonuclear dilepton production data to in-medium
changes of vector meson properties.

Photon-induced dilepton data have the problem that they necessarily
contain a big contribution from the so-called Bethe-Heitler process in
which the incoming photon radiates already a dilepton pair before any
coupling to the hadrons takes place. As is well known, however, this
contribution can be surpressed by proper kinematical cuts
\cite{Effegam,Schaefer}, both for the incoherent and the coherent part
\cite{EffePhD}.

\begin{figure}[ht!]
\centerline{\psfig{figure=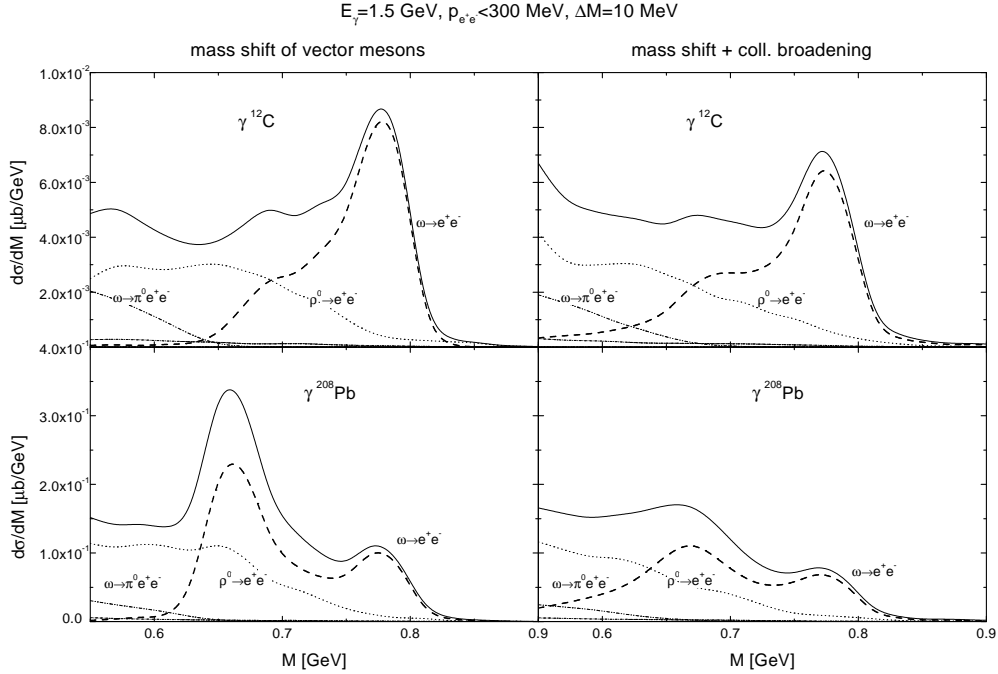,width=14cm}} \caption{\small
Effects of various in-medium modifications of the vector meson
properties on dilepton invariant mass spectra in $\gamma + C$ (top) and
$\gamma + Pb$ (bottom). Only dileptons with a total momentum of $<$ 300
MeV have been considered.} \label{gamdilsens}
\end{figure}

Fig.\ \ref{gamdilsens} then shows the sensitivity of the dilepton
spectra to various in-medium changes of the vector mesons.%
It is immediately apparent, that this sensitivity is just as large as
for the ultrarelativistic heavy-ion collisions. In particular, again
the $\omega$ properties are reflected quite clearly in the (still to be
measured) dilepton spectra. As in the case of pions one can enrich the
in-medium sensitivity by making a cut on the dilepton cm momentum.

\section{Photoabsorption}

All the in-medium effects on vector meson spectral functions are
connected with a significant broadening and corresponding shift of
strength to lower masses. This has an interesting consequence for the
decay of the N(1520) resonance that I have already pointed out in
\cite{MoselHirsch} in 1997. According to the particle data group
listing, which is based on Manley's analysis \cite{Manley} in this
respect, this $D_{13}$ resonance decays on resonance with a probability
of 15 - 25 \% into $N\rho$ and this, even though this decay is
energetically closed if one works with the pole mass of the (broad)
$\rho$. Thus the decay can proceed only through the low-mass tails of
the $\rho$-meson spectral function; the rather large partial decay
width in turn indicates a large coupling constant for the vertex
$N(1520)N\rho$. If the $\rho$ meson's spectral function in nuclear
matter indeed gains strength at low masses then the partial $\rho$
decay width of the $N(1520)$ resonance would increase dramatically.

This effect may already have been experimentally seen. The data for
photoabsorption on nuclei exhibit a universal behavior that --
beginning with rather small mass numbers $A$ -- scales with $A$. This
universal cross section exhibits the $\Delta$ resonance -- though
somewhat broadened --, but the second and third resonance regions
around 1500 and 1650 MeV, respectively, that are well seen in
photoabsorption on the nucleon, have disappeared in nuclei
\cite{Ahrens}.

Fermi-motion alone is the more effective in smearing out resonance
strength the higher the energy of the resonance is. This alone explains
the absence of the third resonance region \cite{Lehr}. The second
resonance region, around 1500 MeV, however, survives the effects of
Fermi-smearing. We have shown in \cite{MoselHirsch,Erice,Lehr} that
this finds a natural explanation in the opening of the $\rho$-decay
channel of the $N(1520)$ resonance. With $\rho$ strength moving in
medium down to lower masses the formerly nearly closed decay opens up
dramatically thus leading to a very large (several hundred MeV) total
width of this resonance. Crucial for this argument is the high partial
decay width of the $D_{13}$ resonance into the $N\rho$ channel. This
has so far, not been directly verified by experiment; the cited large
partial decay width is based only on a theoretical coupled channel
analysis \cite{Manley}. Although it is gratisfying to see that a
similar result also emerges now from the CC calculations of Friman,
Lutz and Wolf \cite{Friman}, a direct experimental verification of this
decay branch would necessitate a detailed partial-wave analysis of
($\gamma,2 \pi$) reactions on the nucleon which so far does not exist.
It is, however, encouraging, that the DAPHNE collaboration
\cite{Daphne} and the TAPS collaboration \cite{Metag}, both at MAMI,
have found experimental evidence for pion invariant mass spectra in
such reactions that show clear deviations from phase-space.

\section{Summary}

In this talk I have shown that pion- and photon-induced reactions show
a sensitivity to in-medium changes of hadrons that is as large as that
observed in ultrarelativistic heavy ion collisions. This result was
originally unexpected because of the much lower baryon densities probed in
such elementary reactions on nuclei ($<1$), compared with those obtained in
ultrarelativistic heavy-ion collisions ($>5$). It finds its explanation
in the fact that all experimentally observed dilepton spectra contain
an integration over the whole reaction history, which, in the case of
heavy-ions, runs from $1 \rho_0$ to more than $5 \rho_0$ and then back
down again to nearly $0 \rho_0$ while the pion- or photon-induced
reaction on nuclei always proceeds close to $\rho_0$, i.e.\ much closer
to equilibrium.

\begin{figure}[ht!]
\centerline{\psfig{figure=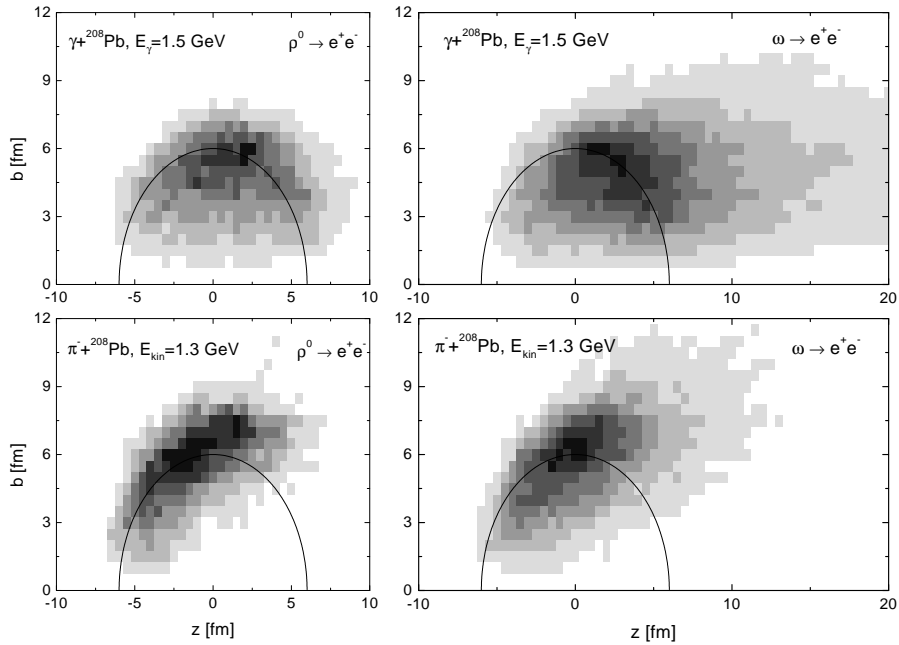,width=13cm}} \caption{\small
Spatial distribution of $\rho \rightarrow e^+e^-$ (left) and $\omega
\rightarrow e^+e^-$ (right) decays in $\gamma Pb$ at 1.5 GeV and $\pi^-
Pb$ at 1.3 GeV.} \label{spacedist}
\end{figure}

In my opinion, an unequivocal identification of the origins of changes
of hadronic spectral functions observed in ultrarelativistic heavy-ion
collisions will not be possible as long as the more elementary
reactions on nuclei have not been studied as well. It is obvious that
hadronic in-medium effects, if they are observed in such reactions at
or below $\rho_0$, have nothing to do with any transition to the QGP
and even a possible connection with chiral symmetry restoration, a very
popular keyword in this field, will be hard to establish (of course,
duality arguments can always be used to keep this connection alive!).
What we can learn, however, are interaction rates of hadrons with
baryons, also for unstable hadrons, and this alone will be an exciting
prospect. Experiments with HADES using the pion beam at GSI and
experiments with photon beams, starting now with TAPS at MAMI, will
help us to clarify the issue of in-medium changes of hadrons. For
example, a close comparison of dilepton spectra obtained in
pion-induced reactions with those obtained with photons should allow
one to sample quite different density regions and thus in-medium
effects, in particular for the $\omega$ meson. This is illustrated
without further words in Fig.\ \ref{spacedist}.
\vspace{1.0cm}

This talk is mainly based on the PhD thesis work of Martin Effenberger
to whom I am grateful for many years of exciting collaboration. The
work has been supported by BMBF, DFG and GSI Darmstadt.

\end{document}